# Effects of interplay of nanoparticles, surfactants and base fluid on the interfacial tension of nanocolloids

A R Harikrishnan[1, #], Purbarun Dhar [2, $], Prabhat K Agnihotri [2], Sateesh Gedupudi [1, &]

and Sarit K. Das[1,2,*]

[1]Department of Mechanical Engineering, Indian Institute of Technology Madras,

Chennai–600036, India

[2]Department of Mechanical Engineering, Indian Institute of Technology Ropar,

Rupnagar–140001, India

[#] Electronic mail: harianilakkad@gmail.com

[$] Electronic mail: purbarun@iitrpr.ac.in

[&]Electronic mail: sateeshg@iitm.ac.in

* Corresponding Author: Electronic mail: skdas@iitrpr.ac.in

Phone: +91-1881-242101

## Graphical Abstract

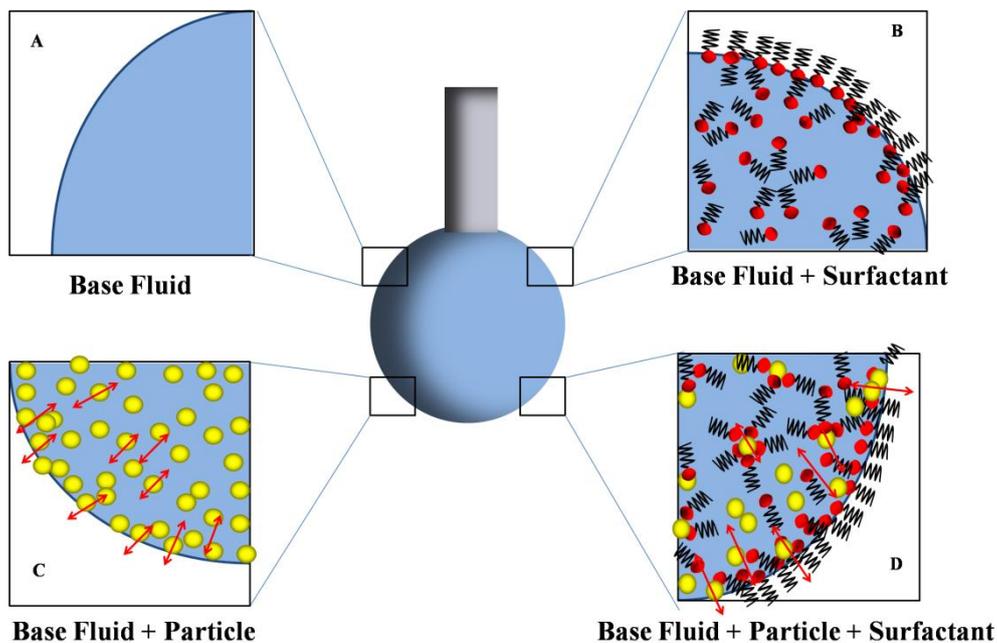

# Abstract

*A systematically designed study has been conducted to understand and clearly demarcate the degree of contribution by the constituting elements to the surface tension of nanocolloids. The effects of elements such as surfactants, particles and the combined effects of these on the interfacial tension of these complex fluids are studied employing pendant drop shape analysis method by fitting Young–Laplace equation. Only particle has shown considerable increase in surface tension with particle concentration in a polar medium like DI water whereas only marginal effect particles on surface tension in weakly polar mediums like glycerol and ethylene glycol. Such behaviour has been attributed to the enhanced desorption of particles to the interface and a mathematical framework has been derived to quantify this. Combined particle and surfactant effect on surface tension of complex nanofluid system showed a decreasing behaviour with respect to the particle and surfactant concentration with a considerably feeble effect of particle concentration. This combined colloidal system recorded a surface tension value below the surface tension of aqueous surfactant system at the same concentration, which is a counterintuitive observation as only particle results in increase in surface tension and only surfactant results in decrease in surface tension. The possible physical mechanism behind such an anomaly happening at the complex fluid air interface has been explained. Detailed analyses based on thermodynamic, mechanical and chemical equilibrium of the constituents and their adsorption-desorption characteristics as extracted from Gibbs adsorption analysis has been provided. The present article conclusively explains several physical phenomena observed, yet hitherto unexplained, in case of interfacial tension of such complex fluids by segregating the individual contributions of each component of the colloidal system.*

## 1. Introduction

Dilute and stable suspensions of nanoparticulate phase in conventional fluids, colloquially termed as nanofluids by the academic community, have revolutionised research in *smart* thermofluidics due to their ability to impart unique properties to the fluid and due to their promise for applications in various micro to macro scale domain. Several research groups have explored the fundamentals of thermophysical properties of these nanosuspensions, such as thermal conductivity and viscosity [1-3]. In spite of the fundamental physical property of surface tension being vital in many engineering applications, it has not been explored much until in the very recent past [4, 5]. Few reports have revealed that the surface tension of such nanosuspensions is grossly governed by the presence of nanostructures, however the roots of the real physics behind this observation is not much explored. Surface tension plays a pivotal role in varied engineering and biological domains such as the wetting dynamics of a surface [5-10], in heat transfer performance of thermal systems such as heat pipes [5], in nucleation and bubble formation and subsequently in phase change heat transfer cases and in determining Critical Heat Fluxes (CHF) in such systems [5], heat exchangers [11], drug interactions with cellular lipid bilayers [12], etc. Moreover, scaling analysis reveals the dominant role played by surface tension forces in case of micro and nanoscale applications [5, 13] and thermofluidic transport, which are among the most sought after technologies in the present day quest for miniaturisation. The past few years have witnessed an increasing trend in the amount of research aimed at understanding the interfacial tension of these complex fluids due to its varied applications. The unanimous opinion of the research community is that the addition of these nanoscale structures are capable of drastically altering the interfacial behaviour [5,7,14] which ultimately provides the capability to tune the interfacial tension so as to optimise the performance of smart devices.

Surfactants are generally used as the stabilising agents in the preparation of nanofluid. The addition of even trace amounts of these surface active agents can drastically affect the interfacial characteristics of the base fluid. Though the studies [6, 15] on the effects of surfactants on the surface tension of nanofluids have been tried, the haphazard planning of the experiments resulted in a lack of clarity and understanding of underlying physical mechanism influencing the surface tension. To the best of the knowledge of the present authors, none of the previous studies have clearly reported the demarcating effects of these surface active agents which influence the interfacial phenomenon. There is an ambiguity with respect to the effect of concentration of the nanoparticle on the behaviour of the interfacial tension of these complex fluids [15, 16]. Few studies [4, 6, 17] report almost a linear increment in surface tension with respect to increase in concentration of solute phase while some authors [5, 10] reported just an opposite trend of decreasing nature, and an increasing and then decreasing behaviour has also been reported [18]. Khaleduzzaman *et al* [15] brief about the lack of clarity on interfacial tension characteristics of nano colloids but could not bring out essential physics behind the problem.

It is clear from the literature review that the nano-scale research community has been unsuccessful in drawing a conclusive picture on the behaviour of the interfacial tension of the nanocolloids with respect to the particle morphology as well as the concentration of solute phase while it has come to a common agreement on the behavioural pattern of the nanofluids with respect to the variations in temperature. The present work tries to bring out the underlying physics of interactions of these nanocolloids and its effect on the nano colloid-gas interfacial tension and hence enabling further exploration and experimentation in different engineering applications where surface tension plays a pivotal role. As a proper analysis and segregation of the different parameters affecting the surface tension is missing in the previous studies, the present study is systematically planned out to dig out the compositional and morphological parameters influencing the interfacial tension characteristics individually and collectively. The compositional parameters of a nano colloid include nature of base fluids, type and concentration of nano particles, concentration and ionic nature of surfactants which are used as stabilisation agents in preparation of nanofluids. The size and the shape of the nanoparticles dictate the morphological characteristics. The current paper deals with the surface tension of colloidal solutions of different metal oxide nanoparticles and deals with the various parameters such as effect of concentration of nanoparticles, nature of the base fluid, morphological characteristics of the particle phase and nature and concentration of the surfactants. The present article explores the physics and mechanisms of the interactions at the interface at the nanoscale and provides in depth analyses to quantify the contributions of thermodynamic, mechanical and chemical mechanisms to the equilibrium interfacial tension of such complex, multi component fluidic systems. To the best of the authors' knowledge, for the first time the contributing effects have been segregated and their individual contributions have been comprehended.

## 2. Materials and methodology

### 2.1. Experimental materials

The experiments for the present article were planned to clearly demarcate and segregate the contributions of each parameter influencing surface tension of nanosuspensions, viz. base fluid properties, nanoparticle type and concentration, surfactant nature and concentration and nanoparticle–surfactant interactions. Deionized (DI) water (polar in nature, synthesized in-situ using a Millipore water processing unit), Ethylene Glycol (weakly polar, 99 % pure, procured from Avra Synthesis, India) and Glycerol (weakly polar, from Merck, India) and five types of metal oxide nanoparticles, viz. CuO (30 nm, NanoArc, Alfa Aeser, India and 80 nm, NaBond, China), $Al_2O_3$ (20nm, Nanoshel Inc. USA), $Bi_2O_3$ (20 nm, Alfa Aeser, India), ZnO ( 80nm, purchased from Nanoshel Inc. USA) and MgO (~ 100 nm, Alfa Aeser, India) have been considered in the present investigation. The nano metal oxides are selected in such a way that the morphological characteristics vary over a range of shapes, from spheres to rods to plates or flakes so as to understand the influence of morphology. Figure 1 (a) shows the TEM image of $Al_2O_3$ showing near–spherical particles and Fig. 1(b1), inset, illustrates the TEM image of CuO (80 nm) and Fig 1(b2) shows the HRSEM image of CuO (30 nm) indicating an oblate morphology. The size of the MgO nanoparticles with disk morphology fluctuates between ~100 nm to 150 nm as illustrated in the HRSEM in Fig. 1 (c). The flake like morphology of $Bi_2O_3$ is clearly evident from Fig 1 (d) with flake thickness ranging from ~13 nm to ~16 nm and average flake length of around ~300 nm. Figure 1 (e) shows the hexagonal pillar structure of ZnO with an average dimension of 50 nm as the face width of the pillar. In the present study, sodium dodecyl sulphate (SDS, 99% pure, Sisco Research Labs, India) has been chosen as the anionic surfactant whereas cetyl trimethylammonium bromide (CTAB) (99% pure, Sisco Research Labs, India) and dodecyl trimethylammonium bromide (DTAB) (AR grade, Avra Synthesis, India) have been chosen as the cationic counterparts.

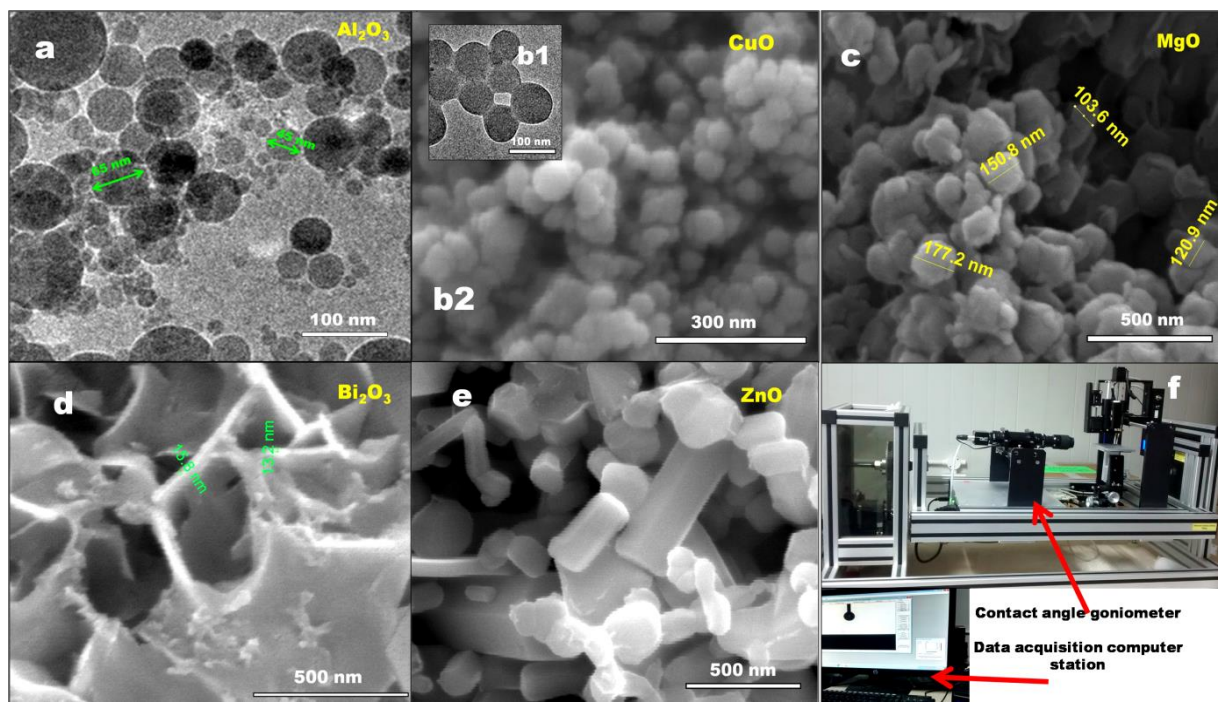

**Figure 1:** (a) TEM image of Al$_2$O$_3$ nanoparticles (~ 20 nm), (b2) HRSEM image of CuO nanoparticles (~ 30 nm) and inset (b1) represents the TEM characterisation of (~ 80 nm)CuO (c)HRSEM characterisation of MgO nanoparticles (~ 100 nm), (d) Bi$_2$O$_3$ HRSEM image showing the flake morphology with flake width ~ 20 nm, (e) ZnO nanostructures with hexagonal rod structure with average dimension ~ 80 nm (f) The experimental and data processing setup.

## 2.2. Experimental methodology

All the experiments were conducted at 30±2 $^\circ$C and humidity of 20±5 %. The complete surface tension study was performed on a standard goniometer (from First Ten Angstroms, model number FTA 1000A, US) by using the principle of pendant drop shape tensiometry. The value of the surface tension is determined by fitting the shape of the suspended pendant drop (from the captured video frames) to the Young-Laplace equation which relates the drop shape to the interfacial tension of the corresponding fluid used. The effective density of the nanosuspensions is measured using a portable density meter (Anton Paar, Germany) and the same is used as in input in the Young-Laplace equation for the determination of surface tension. The inherent advantage of this method over the other commonly used 'Wilhelmy Plate method' is that the cleanliness of the surface does not affect the measurement accuracy and the calibration is done by entering the optical magnification (to match the measured and actual outer diameter of the standard needles used in the dispensing system). Another merit of the pendant method is that it is free from the surface effects that are present in another tensiometric methods such as Wilhelmy plate method and even in du Nouy ring method where there are chances of formation of thin film which brings additional forces such as DLVO forces, particle surface interactions etc. The pendant bubble shape analysis represents the true surface tension especially in case of these complex fluids where foreign particles are dispersed in the base fluid.

The Figure 2 clearly illustrates the classification of experimental protocol followed in the present study so as to clearly demarcate the different parameters contributing to the surface tension. As has been illustrated in the Fig. 2 (a), the first run of experiments have been conducted with only the base fluids such as DI Water, which is inherently polar in nature, ethylene glycol and Glycerol, which are weakly polar solvents, so that the datum line can be fixed which gives the base value of surface tension for comparison. The effect of base fluid on surface tension is referred to as '*A effect*'. The role of surfactants (as shown in Fig 2 (b)), often referred to as the "surface active agents" due to its strong interfacial effects because of the preferential adsorption to the interfaces, has been investigated by employing anionic as well as cationic surfactants. Aqueous solutions of SDS, highly anionic surfactants, with molecular weight 288.4, are prepared and tested for the surface tension over a concentration range varying from 1mM to 20 mM. Similarly for CTAB (molecular weight 364.45) and DTAB (molecular weight 308.34), both cationic in nature, are prepared at different concentration levels ranging from 0.1mM to 2 mM and 2mM to 30 mM and the interfacial tension (IFT) values are measured. From the plot of IFT against the concentration of surfactants, the Critical Micelle Concentration (CMC) of SDS, CTAB and DTAB are observed to be 1mM, 8.5 mM and 15mM which are in close proximity (with a maximum deviation of 10%) to the previously reported values[19] and this validates the present experimental procedure and methods. Hereafter in the present work, the concentrations of surfactant solutions are referred to in terms of CMC as 0.25 CMC, 0.5 CMC etc. based on the above observations. The second set of experimental protocol have been performed to understand the effect of surfactants (*B effect*) on interfacial tension over a wide range of concentrations from 0.25 CMC to 2 CMC.

In order to trace contributions induced by particles due to the adsorption and desorption of these nanoparticles at interface (Fig 2 (c)), referred to as '*C effect*', experiments have been conducted by preparing the nanofluids without using surfactants. The inherent limitation imposed in preparing the nanofluids without surfactants restricted our study to few nanoparticles namely CuO, $Al_2O_3$, ZnO and $Bi_2O_3$ due to the stability issues. CuO and $Bi_2O_3$ were found to exhibit excellent stability without surfactants among the chosen particles even though the other fluids were also stable (relatively very long compared to the timescale of experiments, as determined by sedimentation study by visual inspection). Concentration ranges of nanofluids have been varied from 0.1 wt% to 2.5 wt%. The nanofluids are prepared by adding the weighted amounts of nanoparticles into the base fluid and dispersing the fine particle by sonicating (Oscar Ultrasonics, India) for sufficient amount of time. With the viscous base fluids such as Ethylene Glycol and Glycerol, the nanoparticles are found to be very stable.

The fourth set of experimental protocol has been designed so as to understand the combined effect (*D effect*) of surfactants and nanoparticles in base fluids on interfacial energy of this complex fluid. As illustrated in Fig. 2(d), the fourth set of experimental run is conducted on nanofluids prepared with surfactants which act as stabilising agents in which all the A, B, C effects are present simultaneously and this may have a significant effect at the interface. Apart from these effects, particle – surfactant interaction will also be present which may have an indirect effect on the interfacial phenomenon. The experimental run has been conducted with all the five nanoparticle materials with suitable surfactants. The particle concentration varied from 0.1 wt% to 2.5 wt% and each of the sample have been prepared at three different surfactant concentrations of 0.25 CMC, 0.5 CMC and CMC values of the respective surfactant. Aqueous nanofluids of $Bi_2O_3$ and CuO were found to be stable with both anionic as well as cationic surfactants but whereas for the rest of the nanoparticles only one type of surfactants were used. All the experiments are repeated 3 times in a run and also repeated two times by making solution afresh. The mean values of all the measurements are reported as the experimental observation in the present study with an error that corresponds to the standard deviation of all the measured values.

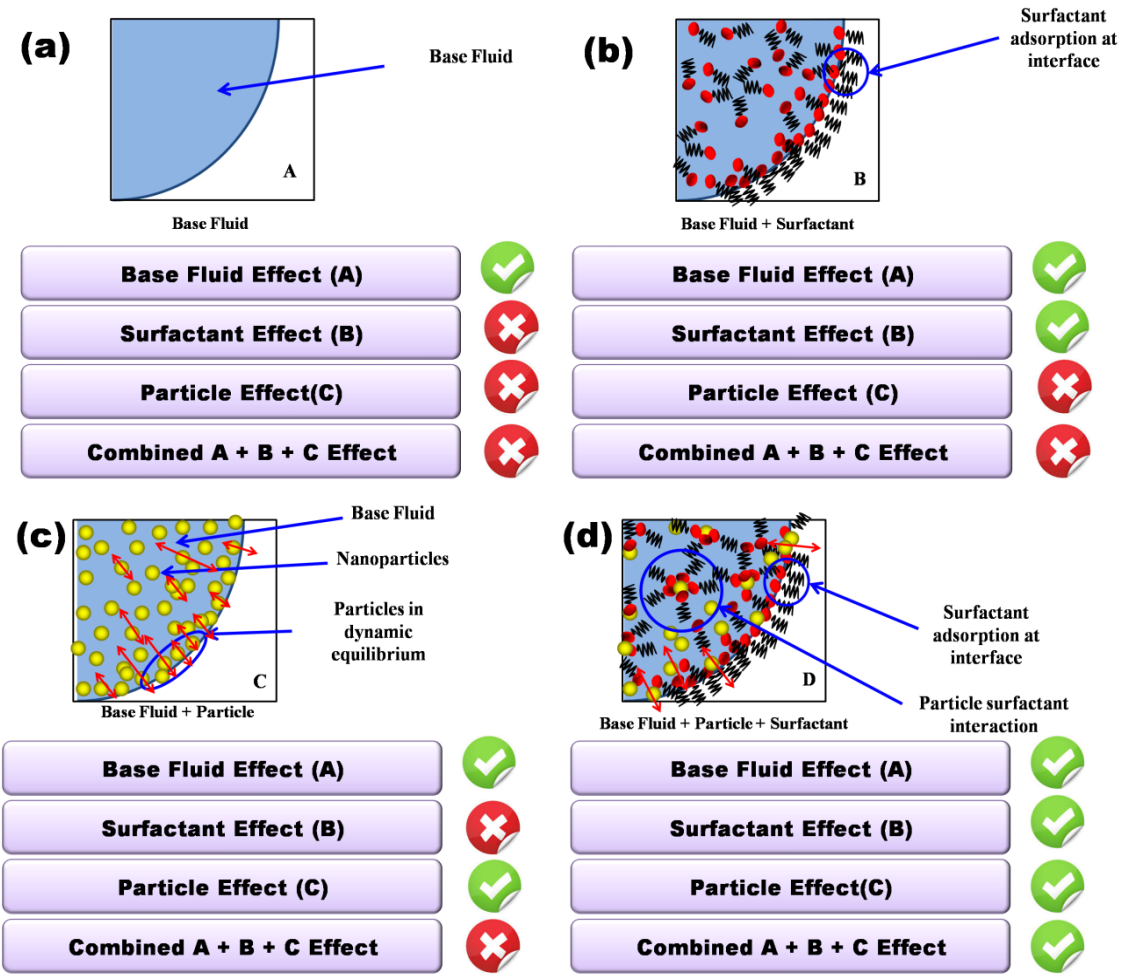

**Figure 2:** Schematic of different experiments to segregate the effect the contributing effects present in each of the protocols.(a) experiments with only base fluid (hereafter referred to as A effect) (b) surface tension studies with only particles (B effect), (c) surface tension experiments on nanofluids with only surfactants (C effect) and (d) surface tension studies on nanofluids prepared with particle and surfactants (D effect).

## 3. Results and Discussions

### 3.1. Base fluid and Surfactants

Initially, surface tension of the base fluids are determined which forms the reference datum and is described as the '*A effect*' as illustrated in Fig 2. The three basefluids which have been considered for the present study are DI water, ethylene glycol (EG) and glycerol (G) with the measured surface tension values being 71.03 ± 0.5 mN/m, 50 ± 0.5 mN/m and 63 ± 0.5 mN/m respectively. The effect of surfactant molecules on the interfacial energy has been referred to as "*B effect*" and the nature of variation of surface tension of aqueous surfactant solution for SDS, CTAB and DTAB is shown in Fig. 3. As mentioned in the previous section, the experiments were conducted over a wide range of concentrations and the CMC value is determined for each surfactant. The surfactant molecules in the solution are governed by the Gibbs adsorption phenomenon and try to form monolayer at the interface which results in the reduction in surface tension of the system. The adsorption characteristics and kinetics of adsorption of these ionic surfactants are well documented in the literature[20-22]. The surfactants dispersed in base fluids such as EG and G could not produce a considerable change in the surface tension even after increasing the surfactant concentrations much higher than the CMC value corresponding to water. The probable reason could be that the surfactant molecules may find it difficult to adsorb to the interface due to the very viscous nature of these base fluids. Viscosity of EG and G are about 10 and 100 times higher than that of water at room temperature, so the diffusion driven surfactant molecules may not be able to overcome the viscous damping, resulting in no change in interfacial energy.

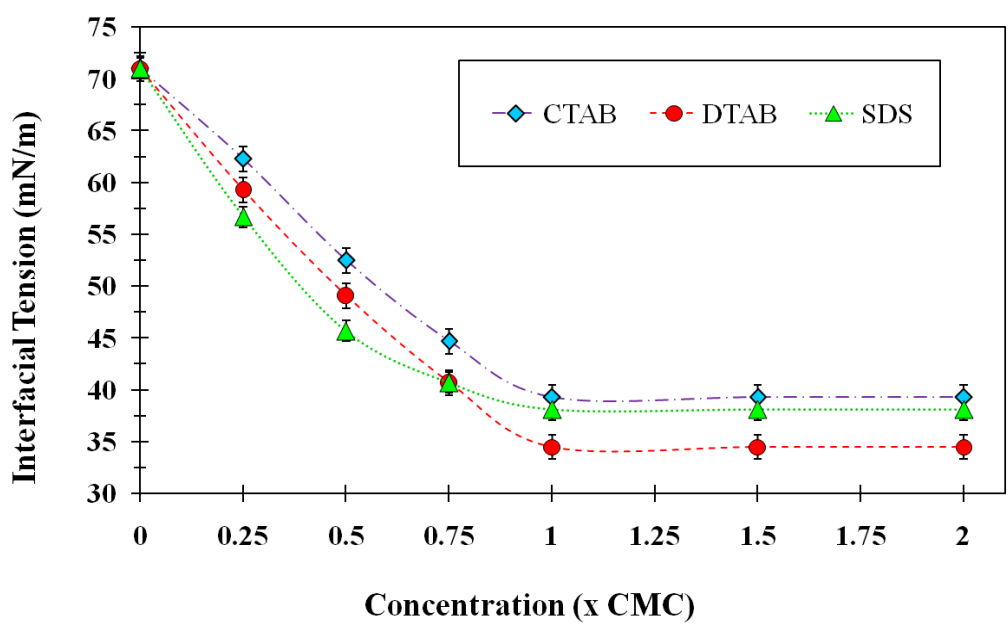

**Figure 3:** Change in interfacial tension of aqueous surfactant solutions with increase in concentration of surfactants (concentrations are expressed in terms of Critical Micelle Concentration (CMC)) for CTAB, DTAB and SDS.

### 3.2. Particle effect

The kinetics of particle adsorption at the interface is a complex phenomenon[14, 23] and the governing physics is controlled by various factors such as particle size, shape of the interface, nature of the base fluid[23], wettability of the particle material at the non–continuum scales[14] etc. Fig. 4 illustrates the effect of different nanoscale particles on the magnitude of the equilibrium interfacial tension of the nanocolloids with change in concentration of the nanoparticle phase. The present investigation considers four particle types, viz. $Al_2O_3$, CuO, ZnO and $Bi_2O_3$ for understanding the effect of particles on the interfacial energy. Among the particles studied, $Bi_2O_3$ exhibits the maximum enhancement of surface tension with respect to the base fluid and $Al_2O_3$ exhibits the minimum increment in the surface energy value. One interesting observation is that CuO (both 30 and 80 nm) and $Al_2O_3$, both having near spherical morphologies, showed considerably minor increment in interfacial energy as compared to that of ZnO and $Bi_2O_3$, which possess hexagonal pillar and flake like morphological characteristics respectively. This reveals that the morphology of the particles as well as the available area to volume ratio for the particle molecules to interact with fluid molecules is also an important aspect that governs interfacial tension in such complex fluids. The nature of variation in values of surface energy of $Al_2O_3$ and CuO particles (both 30 nm and 80 nm) is similar and the values of surface energies are also similar at a particular concentration. The 80 nm CuO nanofluid showed slight increment in the surface tension (about 0.3% - 0.5%) compared to 30 nm nanofluid which is in accordance with previous reports on the size effect of particle on surface tension. $Al_2O_3$ and CuO(s) showed an increase of ~ 3.5% in surface tension at 2.5 wt% compared to the value at 0.1 wt% whereas it is ~ 4% to 5% enhancement for ZnO and $Bi_2O_3$. The inherent limitation of stability posed by the nano suspensions without surfactants limited the study to 2.5 wt%.

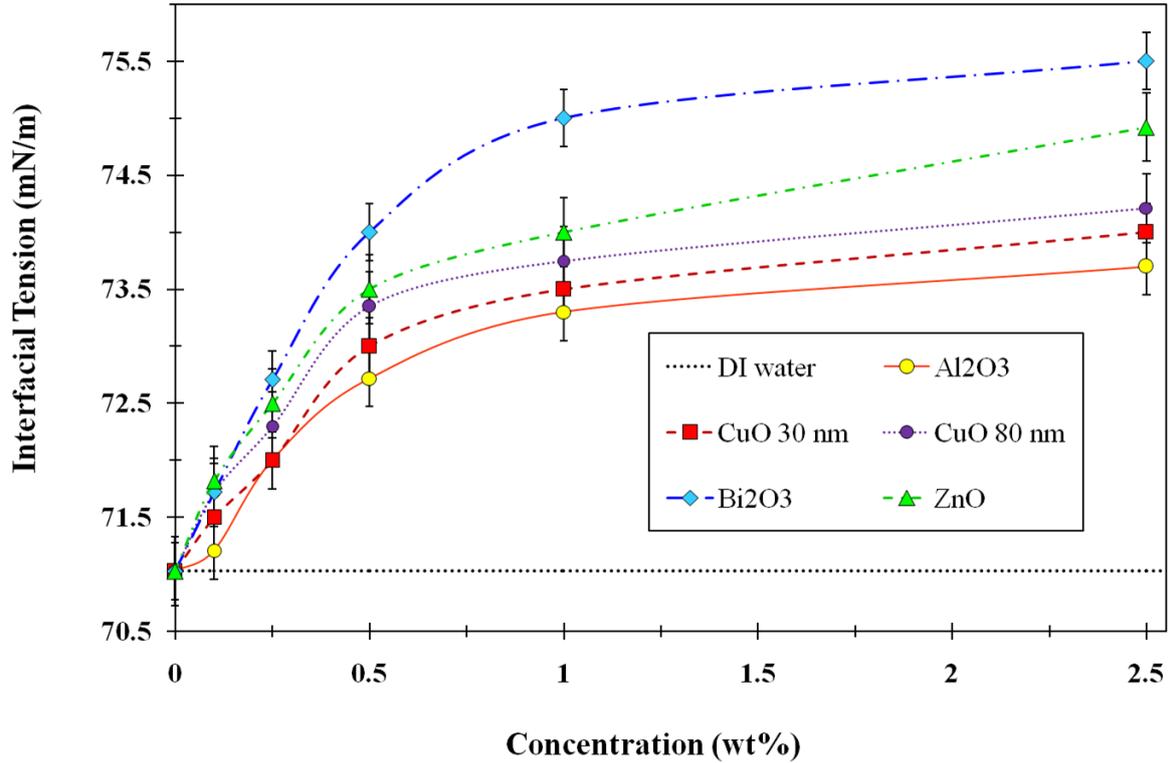

**Figure 4:** Variation of interfacial tension with concentration of nanoparticles in DI water. The black dotted line represents the base reference (surface tension of water that is measured to be 71.03 ± 0.5 mN/m).

Figure 5 illustrates the variation of effective equilibrium surface tension with respect to the concentration of particles in case of EG and Glycerol as base fluids. It is observed that the particles have a marginal effect on the surface energy in case of these two base fluids since the variation with respect to concentration of $Bi_2O_3$ and CuO in these are small and most of the values are within the uncertainty limits of the measurement. The effects of both particles have been studied by conducting the experiments up to concentrations of 5 wt%. Since there are no remarkable effects of particles on these base fluids, further studies on these base fluids are not conducted. The low response can be attributed to the polarity of these fluids, both of which are weakly polar compared to water. Accordingly, the Electric Double Layer (EDL) formed at the particle fluid interface in these cases is thin compared to water. Accordingly, the particles at the interface experience weaker repulsive forces from the particles in the bulk and there is a possible tendency of desorption to the bulk from the interface, leading to no appreciable changes in interfacial tension. Furthermore, because of high viscosity of these base fluids, the particles' Brownian motion might have been partially hindered because of which the particle finds it difficult to adsorb from the bulk to the interface and vice versa.

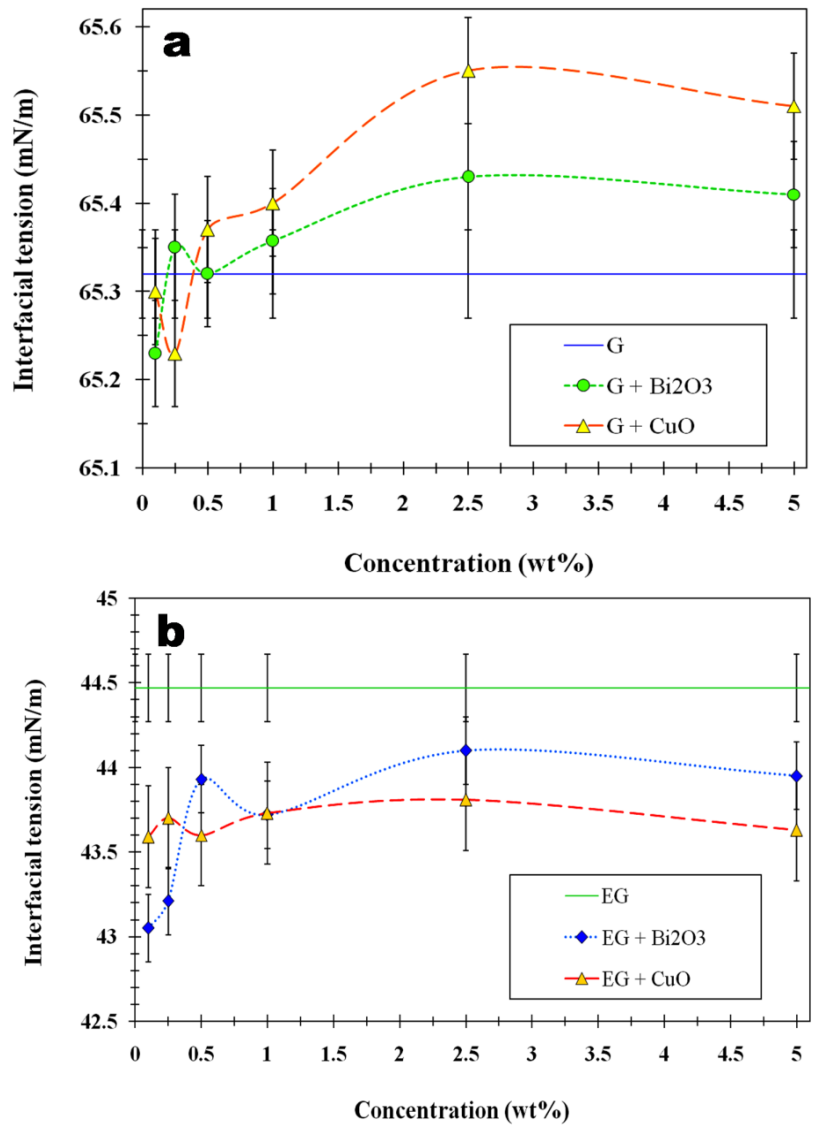

**Figure 5:** Variation of interfacial tension with concentration of nanoparticles in (a) Glycerol (G) and (b) Ethylene Glycol (EG).

The probable governing mechanism behind the increased surface energy behaviour of these complex fluids has been illustrated qualitatively in Fig. 6 (a). The interfacial interaction in case of nano and microsuspensions is very complex and especially when viewed from the nanometre scale, this complex interfacial system is a multiphase zone comprising of interphase of solid nanoparticles, the suspended base fluid and the coexisting interface, which is with atmospheric air in the present study. The surface energy associated with each of these interphases depends on the characteristics of the nanoparticles suspended, the base fluid of the suspension and the surrounding fluid region. The vector addition of these interacting forces at the microscale and summation of all such interactions collectively determine the effective bulk surface tension of the resulting suspension. The affinity of the particles towards the surface, i.e. hydrophilic or hydrophobic nature of the particle, is one of the basic driving factors determining the affinity of the particle towards one of the constituent phases. The change in free energy ($\Delta E_P$) when such a particle (of radius R) moves from the bulk to the fluid–fluid interface in a fluid medium of surface tension γ (assuming the interphase to be planar from the point of view of the nanoscale particle) or vice versa is expressible as[14]:

$$\Delta E_P = -\pi \gamma R^2 (1 + cos\theta)^2 \qquad (1)$$

Accordingly, the particle's affinity towards the surface is a strong function of the equilibrium interphase contact angle and is favoured for the particles that show partial wetting behaviour.

From the point of view of mechanical equilibrium of the particle at the fluid-air interface, the particle must satisfy the minimum energy criteria and will accordingly be positioned at the interface with a height (h) protruding out from the fluid interface (as shown in the Fig. 6a) and creating a localized contact angle of 'θ' (the stable equilibrium contact made by the particle at the nanoscale three phase contact point). Considering the particle radius to be '$r$' (it is noteworthy that a spherical particle assumption has been resorted to in order to simplify the analysis) and the radius of the interfacial area of fluid occupied by the solid particle to be '$b$', the geometrical parameters can be defined as[24]:

$$b = r\, sin\theta \qquad (2)$$

$$h = r(1 - cos\theta) \qquad (3)$$

$$A_1 = 2\pi r h \qquad (4)$$

$$A_2 = 4\pi r^2 - A_1 \qquad (5)$$

where, $A_1$ and $A_2$ are the area of the portion of particle protruding out of the liquid phase and the area of particle submerged in liquid phase respectively. The present analysis assumes the interface to be planar since the radius of curvature of interface between the liquid and gas phase of the pendant droplet is very large compared to the nanoscale particle.

$$e_s = \gamma_{sv}A_1 + \gamma_{sl}A_2 + 2\pi bT \tag{6}$$

$$e_b = \gamma_{sl}(A_1 + A_2) + \pi b^2 \gamma_{lv} \tag{7}$$

where, $e_s$ and $e_b$ represents the energy of the particle when it is present at the interface and when it is within the bulk respectively. The term $\pi b^2 \gamma_{lv}$ represents the energy in submerging the particle and assuming that when a particle is completely in bulk, the modification in liquid vapour surface tension of the fluid is negligible due to absence of any interfacial effects. The symbols $\gamma_{lv}$, $\gamma_{sv}$ and $\gamma_{sl}$ represents the liquid-vapour interfacial tension, solid-vapour interfacial tension and solid-liquid interfacial tension respectively and $T$ is the line tension[24, 25] or energy per unit length of $2\pi b$ units created at the three phase contact of solid-liquid-gas by the nanoparticles. In fact, the traversal of the nanostructures to the liquid–gas interface has also been reported employing molecular dynamics simulations[26].

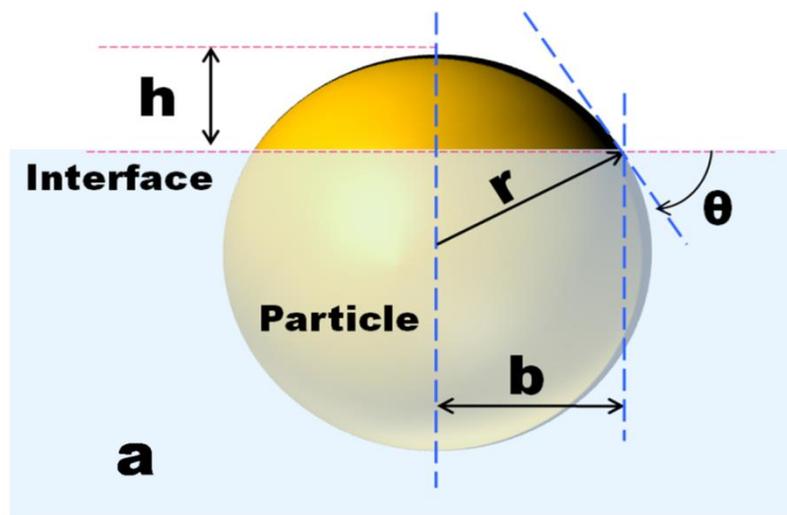

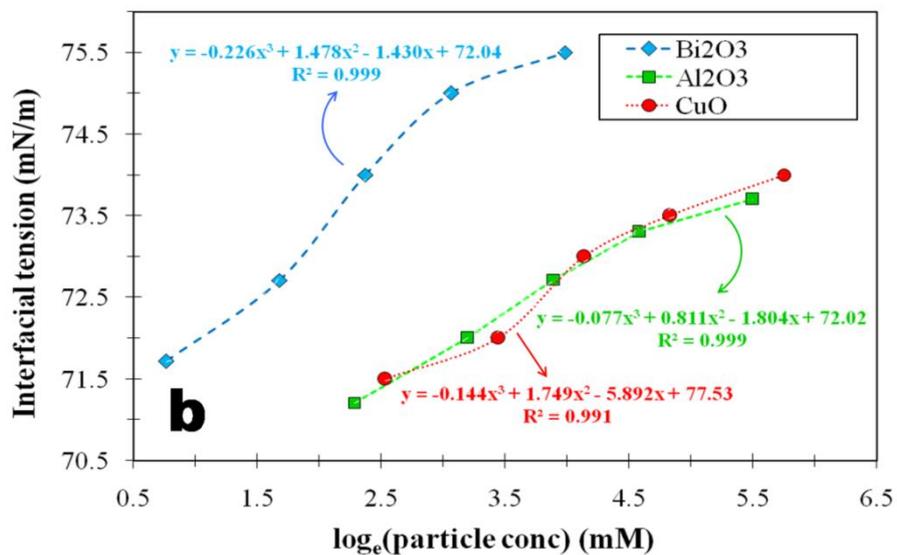

**Figure 6:** (a) Schematic representation of a particle at the fluid-air interface and (b) Gibbs adsorption isotherm polynomial fit for surface tension of only particle based nanofluid with respect to the natural log of concentration of the corresponding particle.

The energy (E) of the interface (liquid–air) created by the base fluid before the addition of particles and the total energy of the complex fluid-air interface ($E_p$) are expressed as

$$E = \gamma_{lv} A_p \tag{8}$$

$$E_p = \gamma_{lv}(A_p - N\pi b^2) + N e_s \tag{9}$$

where, $A_p$ is the area of the pendant drop and $N$ is the total number of particles present at the complex fluid-air interface under equilibrium circumstances. The change in surface tension due to presence of the particles can be expressed as

$$\Delta\gamma = \frac{E_p - E}{A_p} \tag{10}$$

Substituting equations (6), (7), (8) and (9) in equation (10) results in:

$$\Delta\gamma = \frac{N}{A_p}[\,(e_s - e_b) + \gamma_{sl}(A_1 + A_2)\,] \tag{11}$$

It can also be shown that the differential of the energy change with respect to $h$ can be reduced to Eqn. 1 as reported[24] and hence the terms $(e_s - e_b)$ and $\gamma_{sl}(A_1 + A_2)$ are positive which suggests that the change in surface tension of the fluid is positive or alternatively, the consideration of mechanical equilibrium at the interface results in an increase in surface tension of the complex fluid. Of course, the term denoting the difference in energy of the particle at surface and at bulk is positive only in case of particles where potential unfavourable conditions tend to exist for the particle within the bulk.

Considering the thermodynamic aspects of the energetics of particle adsorption at an interface, the free energy change due to adsorption of a nanoscale particle is of the order[23] of ~ $k_B T$ which suggests that the particles are susceptible to thermally excited escape from the interface and are consequently in random motion and are in dynamic equilibrium with the bulk suspension. In the thermodynamic analysis considering the chemical potential of Gibbs adsorption phenomenon, the slope of the plot of surface tension verses the logarithm of concentration of the suspended particle is found to be positive. This suggests that for the particle, desorption away from the interface is observed (as commonly observed in the aqueous solutions of salts). In conjunction to these interacting forces, there are long range forces present between the particle–particle and the particle–base fluid molecules which can also create a considerable shift in the stable equilibrium condition. Electrostatic forces induced due to the formation of electrical double layer around the particle, especially in a strong polar fluid like water, and the electrostatic stress induced deformation of the interface[23] can also affect the equilibrium. The net effect can be the vector addition of the different interaction forces to achieve the stable equilibrium which is a very complex dynamics at the nanoscale. Comparing the results from water as a base fluid on one hand and EG and Glycerol as base fluids on the other hand, it can be observed that the effect of polarity of base fluid, which is indirectly related to the electrostatic stresses and viscous nature of base fluid and which affects the energetics of adsorption/desorption, plays an important role in the prediction of surface tension. This is because the consideration of mechanical equilibrium is applicable to all the base fluids which create an interface with some particles entrapped at the interface of the base fluid – air.

### 3.3. Combined Particle and Surfactant effects

The interaction physics of the nanoparticles and surfactants combination is very complex when it comes to the interfacial phenomenon because of large number of interaction phenomenon at different interphases. The nanofluids which are in real use are prepared with the addition of surfactants considering stability issues. Fig. 7 illustrates the summary of surface tension studies in a matrix form on CuO (30 nm) nanofluids with and without DTAB at various concentrations of nanoparticles and surfactant. A controlled set of experiments with variation of both surfactants and particle concentrations have been carried out in the present investigation. Fig. 8 (a) shows the variation of surface tension of CuO nanofluids prepared with both type of surfactants (anionic and cationic), viz. SDS and DTAB, with change in particle concentration at different concentrations of each surfactant. The CuO nanoparticles were found to be stable for long time periods with both the surfactants and at all three concentrations of surfactants (0.25 CMC, 0.5 CMC and CMC). It can be clearly inferred that as concentration of CuO particle increases at a particular concentration of surfactant, the surface tension is decreasing marginally and as it approaches the higher concentrations the decrease is very feeble or almost nought which suggests that towards higher concentration the particle effect is negligible. This behaviour is found to be similar with both types of surfactants. However, at a given particle concentration, as surfactant concentration increases, there is a considerable decrease in surface tension which indicates that surfactant concentration is having a vital role in determining the surface tension of this complex colloidal system.

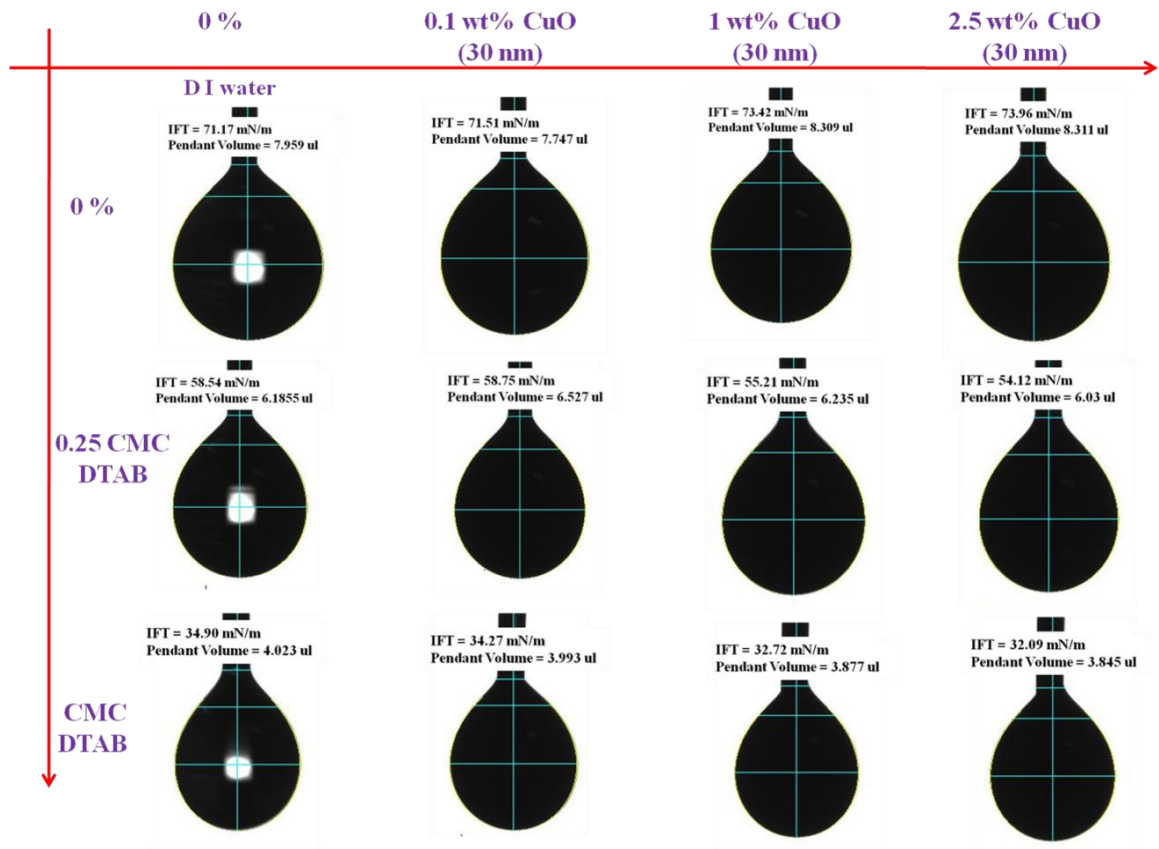

**Figure 7:** Illustration of variation in surface tension of CuO-30 nm (only particle effect), DTAB effect and combined DTAB and CuO at selected particle and surfactant concentration with respect to DI water from the pendant drop shape analysis.

The most important observation is that the surface tension of the complex colloid at a given concentration of particle and surfactant–exhibited a surface tension value lower than the corresponding surface tension of only surfactant solution at the same surfactant concentration. As can be observed from the Fig. 8 (a), even with a small particle concentration of 0.1 wt%, the surface tension was observed to be 54.75 mN/m in case of CuO and SDS combination at 0.25 CMC SDS concentration and 59 mN/m in case of CuO and DTAB combination at 0.25 CMC DTAB concentration which are below the equilibrium surface tension values of 0.25 CMC SDS (56.7 mN/m) and 0.25 CMC DTAB (59.5 mN/m). But the amount of decrease from the equilibrium surface tension of simple surfactant systems varied from surfactant to surfactant and also with surfactant concentration. The above results and those in section 3.2 (particle effect) suggest that the effects of surfactant and nanoparticles on the surface tension of base fluid are not additive in nature when it comes to the combination effect of particle and surfactant in base fluid.

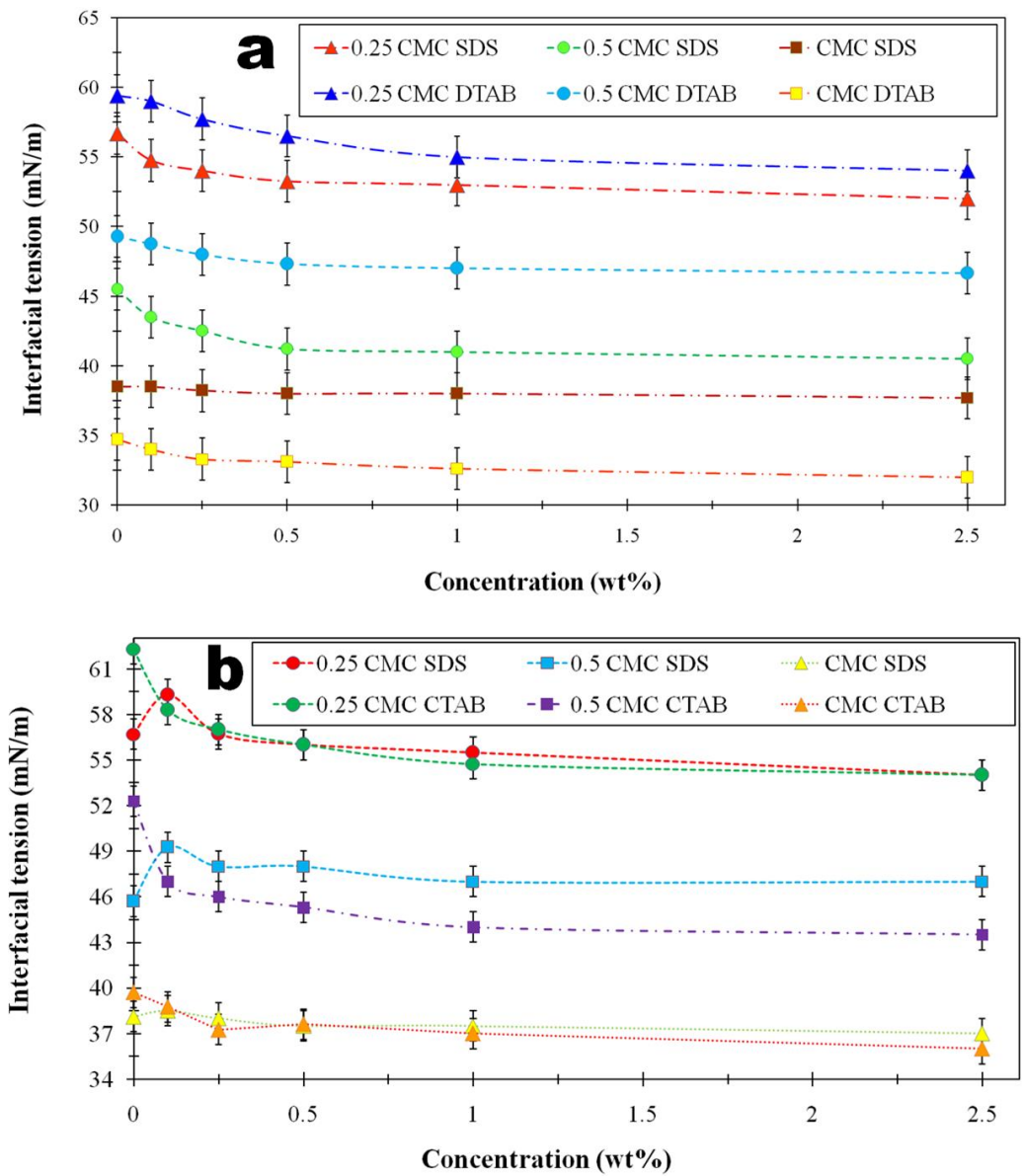

**Figure 8:** Variation of interfacial tension of nanofluids (with surfactants) with concentration of corresponding particle at different surfactant concentrations for (a) CuO with SDS and DTAB and (b) $Bi_2O_3$ with SDS and CTAB.

Similarly, the results of the experiments with nanofluids from combination of $Bi_2O_3$ and SDS and $Bi_2O_3$ and CTAB are illustrated in Fig. 8 (b). The general behaviour of variation in surface tension of complex colloidal system of $Bi_2O_3$ with different surfactant combination is found to be similar to that of the CuO – surfactant colloidal system with respect to concentration of particle and surfactant. However, at lower particle concentration, $Bi_2O_3$ and SDS combination showed a higher value of equilibrium surface tension compared to the aqueous SDS solution at same concentration of surfactant. The effective surface tension of combined system decreased with increase in particle concentration. This change gets negligible towards higher concentration. However, in case of $Bi_2O_3$ and CTAB combination, the effective surface tension is lower than the aqueous CTAB solution at same concentration of surfactant, similar to CuO and surfactant systems discussed earlier. The surface tension for the combined complex system decreased by 6% at 0.25 CMC SDS concentration and 2% at CMC SDS concentration when $Bi_2O_3$ particle concentration is increased from 0.1 wt% to 2.5 wt%. Similarly the effective equilibrium surface tension decreased by 6% at 0.25 CMC CTAB and 4% at CMC CTAB with increase in particle concentration.

$Al_2O_3$ was found to be stable with CTAB and Fig. 9 (a) shows the surface tension variation of $Al_2O_3$ – CTAB with $Al_2O_3$ concentration for three concentrations of CTAB, viz. 0.25 CMC, 0.5 CMC and at CMC. Similarly Fig. 9 (b) shows the surface tension variation for the MgO and ZnO nanofluids for two different concentrations of surfactant. The nature of variation of surface tension with concentration of particle and surfactant are found to be similar to the above cases of CuO and $Al_2O_3$.

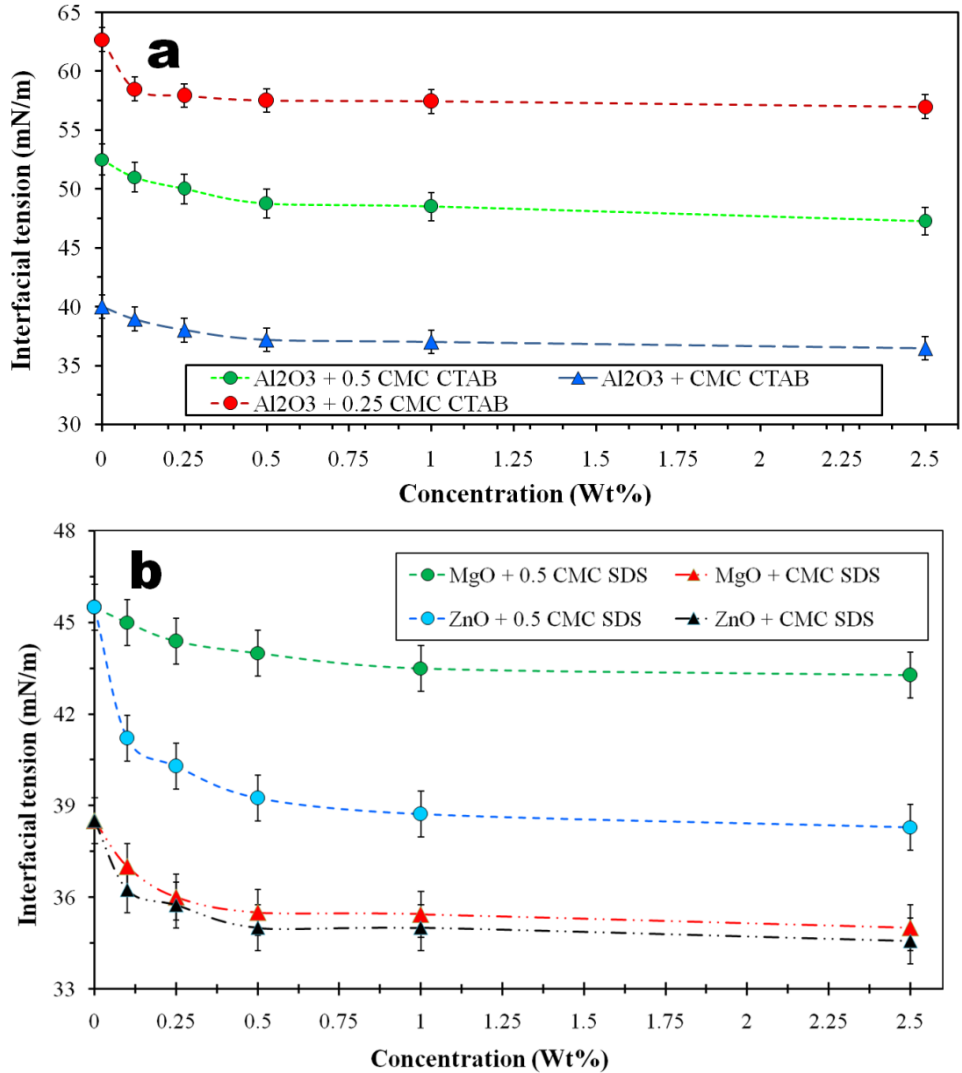

**Figure 9** Variation of interfacial tension of nanofluids (with surfactant) with concentration of corresponding particle at different surfactant concentration for (a) $Al_2O_3$ with three concentrations of CTAB and (b) MgO and ZnO with different concentrations of SDS.

Another crucial observation from Figures 8 and 9 is that the addition of particles has resulted in the modification of CMC values of the resulting complex fluid system, revealed from the fact that the surface tension is found to decrease with the increase in particle concentration in case of systems in which surfactant concentration was maintained at CMC. Once the particles are added to the surfactant solution to form the nanofluid, surfactant molecules adsorb to the particle at the solid–liquid interfaces which results in the alteration of effective CMC value of the resulting fluid system. A quantitative description of each effect, viz. *A, B, C and D effects* discussed earlier is illustrated in Fig. 10.

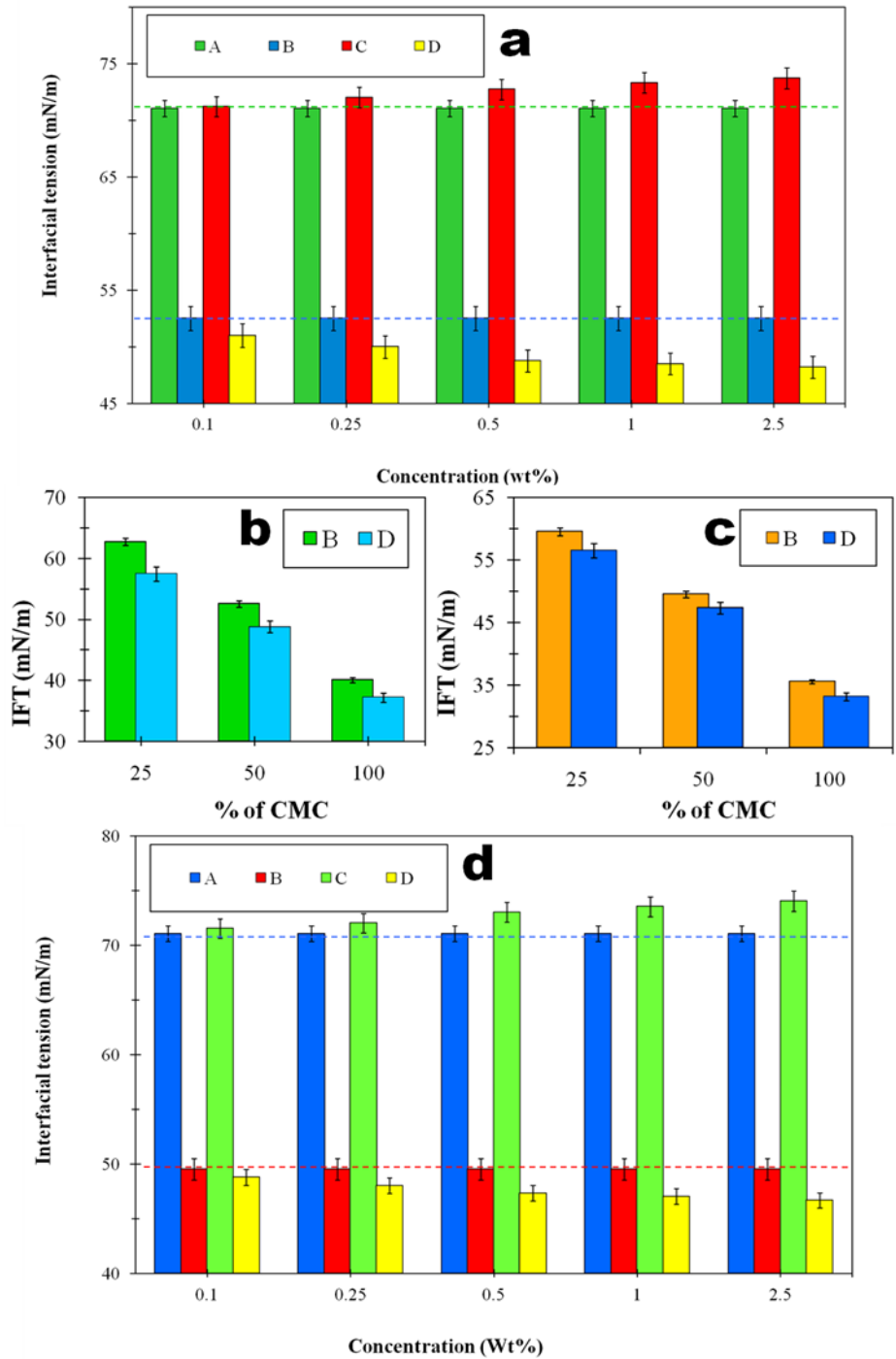

**Figure 10:** Quantitative representation of IFT of base fluid (A effect), surfactant solution (B effect), particle in base fluid (C effect) and particle and surfactant combination (D effect) for (a) $Al_2O_3$ at different particle concentration and at CTAB concentration of 0.5 CMC (b) 0.5 wt. % particle concentration $Al_2O_3$ at different CTAB concentrations (c) 0.5 wt. % CuO (30 nm) nanofluid at different DTAB concentration and (d) CuO (30 nm) at different particle concentration at DTAB concentration of 0.5 CMC.

The probable physical mechanism behind the anomalous behaviour of the combined surfactant particle effect can be explained based on adsorption kinetics. In a normal aqueous surfactant solution, a fraction of surfactant molecules adsorb at the interface according to the Gibbs adsorption isotherm resulting in the decrease in surface tension of the base fluid. The surface excess concentration ($\Gamma$) depends on many factors such as temperature, nature of surfactant molecule, its activity coefficient, concentration etc. as per the adsorption equation given by Gibbs[27, 28]:

$$\Gamma = \frac{-1}{RT}\frac{d\gamma}{d(\ln(a))} = \frac{-1}{2RT}\frac{d\gamma}{d(\ln(c))} \quad (12)$$

where '$R$' is the universal gas constant, '$T$' is the absolute temperature and '$a$' is the activity. In case of a univalent electrolyte and in dilute suspensions, the mean activity coefficient can be approximated to be equal to one so that the Gibbs adsorption equation can be expressed in terms of concentration in solution, '$c$'.

Once the particles are added to the fluid with surfactant molecules, some portion of the total surfactant molecules tries to adsorb at the solid-liquid interface of particle and base fluid and this depends on the nature of particle and base fluid. Another fraction of molecules will try to adsorb at the base fluid-air interface. A spherical cap of surfactant molecules adsorbed to the particle will help to stabilise the suspension. The presence of a foreign nanoparticle in the aqueous surfactant solution drastically modifies the localized adsorption characteristics because of the creation of one more interfacial layer between numerous nanoparticles surfaces and the base fluid which constitute preferential adsorption sites. As explained in the previous section, the nanoparticles at the gas-liquid interface are in dynamic equilibrium with the bulk because of low energy change required for adsorption. Assuming that the surfactant capped nanoparticle's energetics of adsorption to be of the same order of magnitude as that of only particle, the former will also be in dynamic equilibrium state. This results in nanoparticle driven surfactant adsorption at the liquid-air interface as shown in the Fig. 11 (a), with the hydrophobic tail of the surfactant in the gas phase. The resulting effect is that more number of surfactant molecules is driven towards the liquid–air interface, which results in further decrease of surface tension compared to the aqueous surfactant solution. Since the particle energetics of adsorption-desorption is less and the natural tendency of surfactant to get adsorbed at the liquid–air interface is more, at some instant of time both the driving potential will be aiding and during other instants of time, it will be opposing. The dynamicity tries to bring the complex colloidal system to a state of dynamic equilibrium and this depends on the particle and surfactant. In the aiding phase, the nanoparticles act as a carrier of surfactant molecules to the interface which helps to populate the liquid–air interface with more surfactant molecules. In case of combined surfactant–particle combination, surfactant molecules will be adsorbed to most of the nanoparticles which eliminate the particle alone getting adsorbed to the interface. So, almost all the adsorbate at the liquid -air interface consists of surfactant molecules.

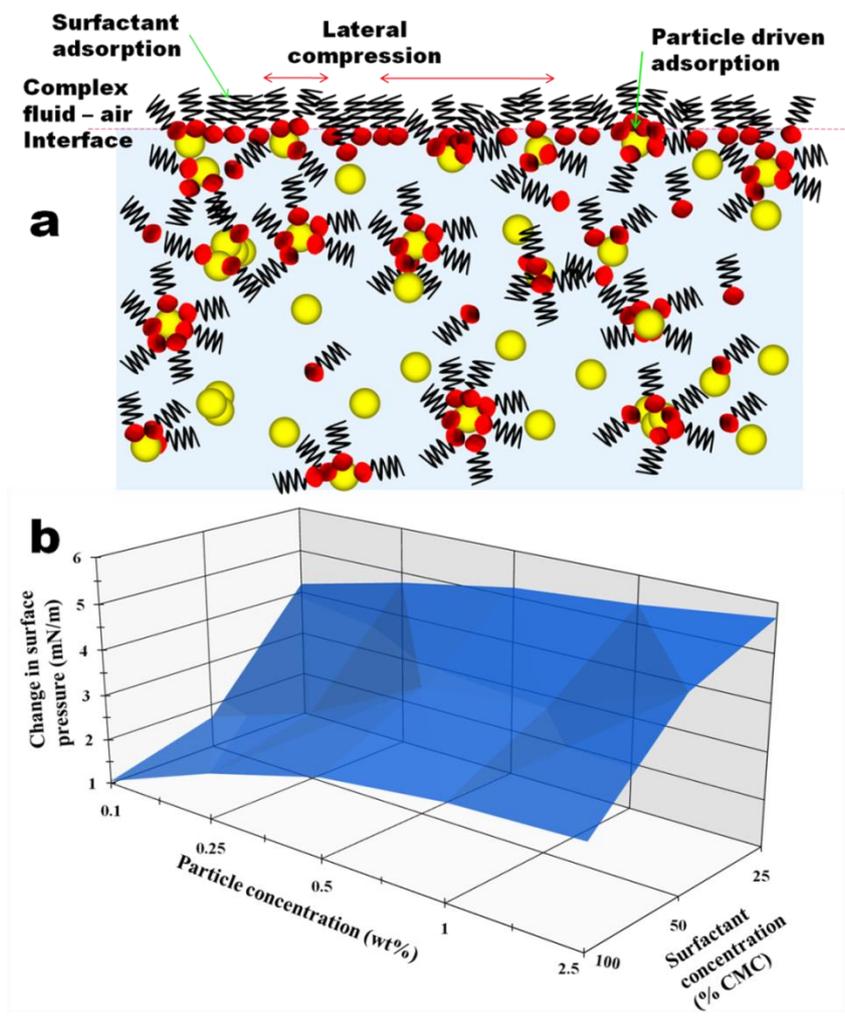

**Figure 11:**(a) Illustration of a physical mechanistic possible model explaining the interfacial interactions which affect surface tension in case of combined surfactant – nanoparticles combination (b) Surface plot illustrating the change in surface pressure with respect to the change in concentration of nanoparticles and concentration of surfactant.

As observed in Figures 8, 9 and 10, with increase in particle concentration, the effective equilibrium surface tension of complex fluid exhibits a considerable difference only during the initial phase and towards higher concentration and the decrease is marginal or tending to zero in some cases. As the particle concentration increases initially, it carries more surfactant to the base fluid–air interface which results in decrease of surface tension but as particle concentration increases, the increased surfactant concentration at the interface (carried by the increased particle population) results in steric hindrance among adsorbed surfactant molecules which results in a marginal decrease or near constant value of surface tension. When the available area of the monolayer of adsorbed molecules of surfactant and particles are more, the distance between the adjacent molecules is large and so interaction will be weak. When the surface area is reduced or a compression effect[28] is created due to additional molecular adsorption, the molecules will exert a repulsive force on each other. This effect is expressed in terms of a two dimensional analogue of pressure called as surface pressure ($\pi_s$) as:

$$\pi_s = \gamma_o - \gamma \tag{13}$$

where $\gamma_o$ is the surface tension of the base fluid and $\gamma$ is the surface tension with monolayer.

More insight into this physics can be obtained from the surface pressure change ($\Delta\pi_s$) analysis of the complex fluid interface as illustrated in Fig. 11 (b). The surface plot shown in Fig. 11 (b) illustrates the change in surface pressure with respect to the concentration of nanoparticles as well the concentration of surfactant for $Al_2O_3$ and CTAB combination. In the present analysis, the change in surface pressure ($\Delta\pi_s$) is defined as the surface tension of aqueous surfactant solution minus the surface tension of complex nanofluid with particle and surfactant. Basically, it represents the reduction in surface tension of aqueous surfactant solution with the addition of nanoparticles. As more and more surfactant molecules are getting adsorbed at the liquid–air interface due to particle driven adsorption, the surfactant monolayer is getting compressed which results in decrease of surface tension and thereby an enhancement in surface pressure. It is evident from the surface plot in Fig.11 (b) that as the nanoparticles concentration is increases in the initial phase; the surface pressure also increases and towards higher particle concentration, the hike in the value of $\Delta\pi_s$ gradually dies out due to the steric hindrance. Also, for a given particle concentration, it can be observed that as the surfactant concentration increases, the $\Delta\pi_s$ decreases which may be due to the fact that at lower surfactant concentration, more active adsorption sites are available at the liquid–air interface and as surfactant concentration increases, the active sites decreases since it has already been occupied by surfactant molecules at higher concentration. Moreover, the rate of change of $\Delta\pi_s$ is more drastic at lower particle concentration which again is in accordance with our physical mechanistic model since at lower particle concentration, the number of surfactant molecule carriers (i.e. nanoparticles) is less which results in less transport of surfactant molecules to the complex fluid–air interface. A similar analysis by taking into consideration of surface pressure has been reported elsewhere[29]. The irregular behaviour showing an enhanced surface tension value for combined $Bi_2O_3$ and SDS compared to aqueous SDS solution at the same concentration may be because the SDS is incapable of carrying the $Bi_2O_3$ nanoparticle to the surface which results in lower surfactant surface excess at the liquid–air interface compared to the same concentration of aqueous SDS since some of the SDS will be attached to the $Bi_2O_3$ particle liquid interface.

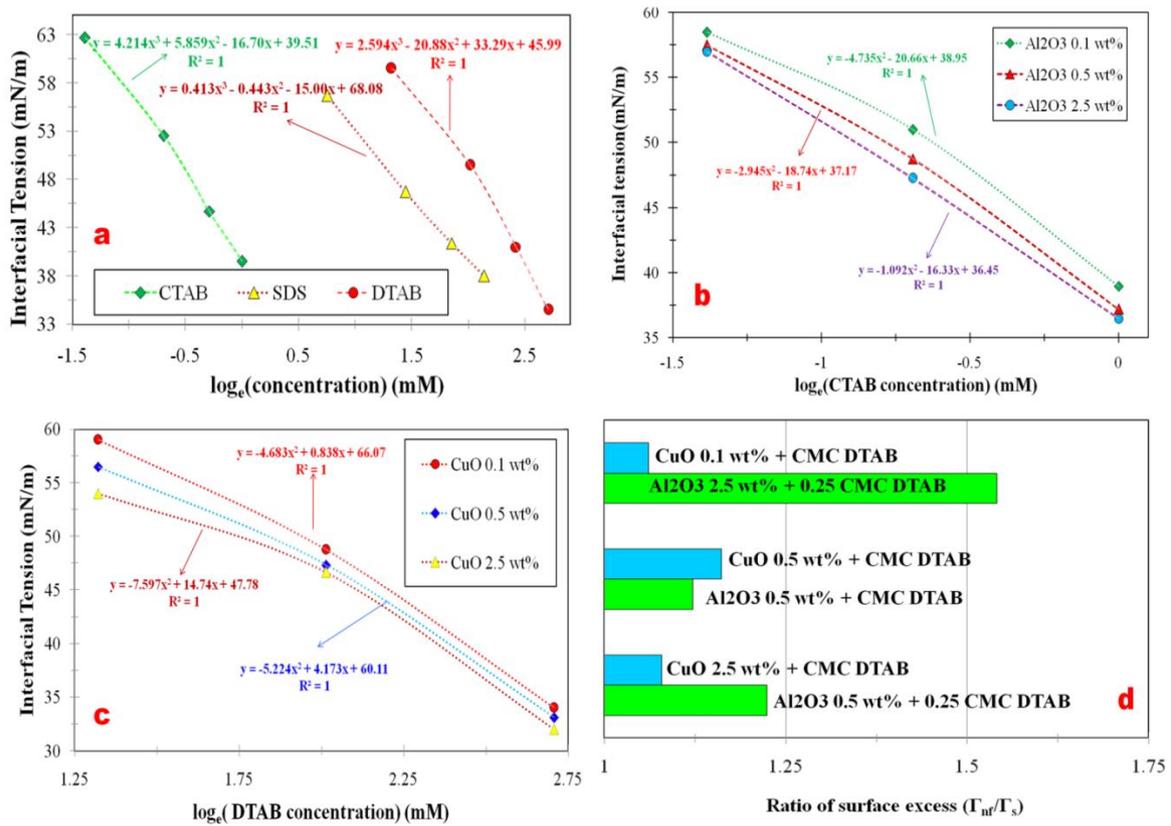

**Figure 12:** Gibbs adsorption isotherm curve fitting (interfacial tension verses natural log of surfactant concentration) for (a) aqueous solutions of SDS, CTAB and DTAB (b) $Al_2O_3$ nanofluids with CTAB surfactants at three particle concentrations (in wt%) of $Al_2O_3$ (c) CuO nanofluids with DTAB at three concentrations of CuO and (d) bar diagram illustrating the ratio of surface excess ($\Gamma_{nf}$) in case of combined surfactant–particle nanofluids to the surface excess in case of aqueous surfactant solution ($\Gamma_s$) for few cases (sample names indicated against corresponding bar).

A better comprehension of the interfacial physics can be obtained from analysis of the associated Gibbs adsorption of surfactants and nanofluids. A quantitative picture of the surface excess of a given species (particles of surfactant molecules) can be drawn from the use of Eqn. (12) and from the plot of surface tension versus the log of concentration as illustrated in Fig.11. Fig. 12(a) illustrates the variation of surface tension of each of the aqueous solutions for each surfactant in the present study with natural logarithm of concentration of surfactant with an appropriate polynomial function fit to obtain the equation of the curve. Comparing Eqn. (12) with the derivative of each adsorption curve, the surface excess ($\Gamma_s$) for a given sample is obtained for a particular concentration of surfactant as

$$\Gamma|_{c=c^*} = \frac{-1}{RT}\frac{d\gamma}{d(lna)}|_{c=c^*} = \frac{-1}{2RT}\frac{d\gamma}{d(lnc)}|_{c=c^*} = \frac{-1}{2RT} * (slope\ of\ curve)|_{c=c^*} \quad (14)$$

Hence, the surface excess in each case and at a particular concentration is directly proportional to the slope of the curve at that particular concentration. Fig. 12(b) illustrates the Gibbs adsorption polynomial fitting for $Al_2O_3$ and CTAB combination in water for three nanoparticle concentrations of 0.1, 0.5 and 2.5 wt%. Assuming that the Gibbs adsorption equation is equally valid in case of surfactant particle combination as is valid for surfactant case; the same analysis can be applied to the surfactant stabilized nanofluids to obtain the surface excess in case of such nanofluids ($\Gamma_{nf}$). The same analysis has been extended to the particle and surfactant combination of CuO and DTAB and illustrated in Fig.12 (c). The ratio (Eqn. 14) of surface excess of stabilized nanofluid to that of the aqueous surfactant solution for a given surfactant provides a quantitative analysis of the species population contributing to the surface excess.

$$\frac{\Gamma_{nf}|_{c=c^*}}{\Gamma_s|_{c=c^*}} = \frac{(slope\ of\ curve)_{nf}|_{c=c^*}}{(slope\ of\ curve)_s|_{c=c^*}} \quad (15)$$

Similar analysis extended to different particle-surfactant combination shows that the surface excess ratio ($\Gamma_{nf}/\Gamma_s$) to be greater than unity, indicating an increase in surface excess in case of surfactant stabilized nanofluids. As the surface excess is higher in case of particle surfactant combined case, it justifies the physical mechanism of enhanced surfactant presence, as explained earlier, which results in a reduced magnitude of effective equilibrium surface tension. Fig. 12 (d) illustrates the surface excess ratio for some of the combined cases of $Al_2O_3$ and CuO. In case of $Al_2O_3$ and CTAB (CTAB at 0.25 CMC), as the concentration of particle increases from 0.5 wt% to 2.5 wt%, the surface excess ratio increases which could possibly be due to the increased particle population acting as transporters of the surfactant molecules to the interface. Whereas in case of the same particle concentration of 0.5 wt% $Al_2O_3$, as the surfactant concentration increases from 0.25 CMC to CMC, the surface excess ratio decreases because of the excess amount of surfactant already present at the liquid air interface (transported by the existing particle phase) which hinders any further adsorption. In case of CuO and DTAB combination (keeping the DTAB concentration constant at CMC), a general trend in the variation of surface excess ratio with particle concentration could not be arrived at. This suggests that the adsorption dynamics in case of combined particle–surfactant combination is a strong function of the nature of surfactant and particle and of their interactions in a polar liquid phase. A few cases also showed a lower value (maximum of 9% reduction) of surface excess in case of combined particle surfactant combination in comparison to aqueous surfactant case. These may be partly due to the experimental uncertainties and presence of impurities in the particle phase and partly due to degree of hindered (thermally or mechanically during experiments) interplay of interaction among the particle, surfactant molecules and the fluid phase.

## 4. Conclusion

A systematically organised study has been conducted to understand the effect of surfactants, nanoparticles and the combined effect of surfactant and nanoparticles on the interfacial tension of such complex fluids. The present approach utilises the principle of pendant droplet method due to its inherent advantages, especially in case of complex fluid systems such as nanofluids where there are several interaction forces which govern the surface tension. Only particles in base fluid showed an enhancement in surface tension with increase in nanoparticles concentration. The establishment of a complete thermodynamic equilibrium criterion is found to be the possible governing mechanism and it constitutes the balance between mechanical, chemical, thermal and electrostatic driving potentials. Explanations of the governing forces at the interfaces and their influence vis-à-vis interfacial adsorption–desorption behaviour of nanosized particles have been provided in the present article. Combined particle and surfactant combination was found to exhibit a surface tension value below the aqueous surfactant case and showed a marginal decrease with increasing particle concentration for a given surfactant concentration. In case of combined particle surfactant case, the effective surface tension is not additive of surfactant effect and particle effect and this anomaly has been explained based on Gibbs adsorption theory by quantifying the propensity of absorption in such systems. The present article conclusively sheds insight into several issues related to the interfacial tension of complex multicomponent nanofluid systems, both qualitatively and quantitatively.

## Acknowledgement

The authors would like to thank Dr. HimanshuTyagi of IIT Ropar for the ultrasonicator facility. ARH would like to thank Ministry of Human Resource Development, Govt. of India, for the doctoral scholarship.

## References

**1.** Das, S. K; Choi, S. U. S; Patel, H. E. Heat transfer in nanofluids—A Review. *Heat Transfer Eng.* **2006**, 27, 3-19.

**2.** Kumar, D. H; Patel, H. E; Kumar, V. R. R; Sundararajan, T; Pradeep, T; Das, S. K. Model for heat conduction in nanofluids. *Phys. Rev. Lett.* **2004**, 93, 144301.

**3.** Khanafer, K; Vafai, K. A critical synthesis of thermophysical characteristics of nanofluids. *Int. J. of Heat and Mass Transfer.* **2011**, 54, 4410-4428.

**4.** Bhuiyan, M. H. U; Saidur, R; Mostafizur, R. M; Mahbubul, I. M; Amalina, M. A. Experimental investigation on surface tension of metal oxide–water nanofluids. *Int. Comm. Heat and Mass Transfer.* **2015**, 65, 82-88.

**5.** Chinnam, J; Das, D. K; Vajjha, R. S; Satti, J. R. Measurements of the surface tension of nanofluids and development of a new correlation. *Int. J. of Thermal Sciences.* **2015**, 98, 68-80.

**6.** Tanvir, S; Qiao, L. Surface tension of nanofluid-type fuels containing suspended nanomaterials. *Nanoscale res.lett.* **2012**, 7, 1.

**7.** Vafaei, S; Wen, D; Borca-Tasciuc, T. Nanofluid surface wettability through asymptotic contact angle. *Langmuir.* **2011**, 27, 2211-2218.

**8.** Chen, Ruey-Hung, Tran X. Phuoc, and Donald Martello. "Surface tension of evaporating nanofluid droplets." *Int.J. of Heat and Mass Transfer.* **2011**, 54, 2459-2466.

**9.** Lim, S; Horiuchi, H; Nikolov, A. D; Wasan, D. Nanofluids alter the surface wettability of solids. *Langmuir.* **2015**, 31, 5827-5835.

**10.** Waghmare, P. R; Mitra, S. K. Contact angle hysteresis of microbeadsuspensions. *Langmuir.* **2010**, 26, 17082-17089.

**11.** Pantzali, M. N; Kanaris, A. G; Antoniadis, K. D; Mouza, A. A; Paras, S. V. Effect of nanofluids on the performance of a miniature plate heat exchanger with modulated surface. *Int. J. of Heat and Fluid Flow.* **2009**, 30, 691-699.

**12.** Feller, S. E; Pastor, R. W. Constant surface tension simulations of lipid bilayers: the sensitivity of surface areas and compressibilities. *The J. of chem. Phys.* **1999**, 111, 1281-1287.
**13.** Kandlikar, S. G. Scale effects on flow boiling heat transfer in microchannels: A fundamental perspective." *Int. J. of Them. Sci.* **2010**, 49, 1073-1085.
**14.** Ravera, F; Santini, E; Loglio, G; Ferrari, M; Liggieri, L. Effect of nanoparticles on the interfacial properties of liquid/liquid and liquid/air surface layers. *The J. of Phys. Chem. B.* **2006**, 110, 19543-19551.
**15**. Khaleduzzaman, S. S; Mahbubul, I. M; Shahrul, I. M; Saidur, R. Effect of particle concentration, temperature and surfactant on surface tension of nanofluids. *Int. Comm. in Heat and Mass Transfer.* **2013**, 49, 110-114.
**16.** Chaudhuri, R. G; Paria, S. The wettability of PTFE and glass surfaces by nanofluids. *J. of colloid and int. Science.* **2014**, 434, 141-151.
**17.** Huminic, A; Huminic, G; Fleaca, C; Dumitrache, F;Morjan, I. Thermal conductivity, viscosity and surface tension of nanofluids based on FeC nanoparticles. *Powder Tech.* **2015**, 284, 78-84.
**18.** Vafaei, S; Purkayastha, A; Jain, A; Ramanath, G; Borca-Tasciuc, T. The effect of nanoparticles on the liquid–gas surface tension of Bi2Te3 nanofluids. *Nanotech.* **2009**, 20, 185702.
**19.** Aguiar, J; Carpena, P; Molina-Bolıvar, J. A; Ruiz, C. C. On the determination of the critical micelle concentration by the pyrene 1: 3 ratio method. *Journal of Colloid and Interface Science,* **2003**, 258, 116-122.
**20.** Eastoe, J; Dalton, J. S. Dynamic surface tension and adsorption mechanisms of surfactants at the air–water interface. *Advances in colloid and interface science,* **2000**, 85, 103-144.
**21.** Prosser, A. J; Franses, E. I. Adsorption and surface tension of ionic surfactants at the air–water interface: review and evaluation of equilibrium models. *Colloids and Surfaces A: Physicochemical and Engineering Aspects.* **2001**, *178,* 1-40.
**22.** Paria, S; Khilar, K. C. A review on experimental studies of surfactant adsorption at the hydrophilic solid–water interface. *Adv. in colloid and interface science,* **2004**, 110, 75-95.
**23.** Zeng, C;Bissig, H; Dinsmore, A. D. Particles on droplets: From fundamental physics to novel materials. *Solid state comm.* **2006**, 139, 547-556.
**24.** Wi, H. S; Cingarapu, S; Klabunde, K. J; Law, B.M. Nanoparticle adsorption at liquid–vapor surfaces: influence of nanoparticle thermodynamics, wettability, and line tension. *Langmuir,* **2011**, 27, 9979-9984.
**25.** McBride, S. P; Law, B. M. Influence of line tension on spherical colloidal particles at liquid-vapor interfaces. *Phys. Rev. Let.* **2012**, 109, 196101.
**26.** Lu G, Duan Y Y and Wang X D, Surface tension, viscosity and rheology of water-based nanofluids: a microscopic interpretation on the molecular level, *J Nanopart. Res,***2014**, *16 (9),* 1-11.
**27.** Russel, W. B; Saville, D. A; Schowalter, W. R. "Colloidal dispersions", *Cambridge University Press.* **1989**, (ISBN 0 521 34188 4).
**28.** Ghosh, P. "Colloid and Interface Science", *PHI Learning*, New Delhi, **2009**.

29. Garbin, V; Jenkins, I; Sinno, T; Crocker, J. C; Stebe, K. J. Interactions and Stress Relaxation in Monolayers of Soft Nanoparticles at Fluid-Fluid Interfaces. *Phys. Rev. Let.* **2015**, 114, 108301.